# Relational Archetypes: A Comparative Analysis of AV-Human and Agent-Human Interactions

Antoni Lorente[1], Amin Oueslati[1], Robin Staes-Polet[1]

[1]The Future society

**Abstract**

Over the last couple of years, AI Agents have gained significant traction due to substantial progress in the capabilities of underlying General Purpose AI (GPAI) models, enhanced scaffolding techniques, and the promise to drive societal transformation. Companies, researchers, and policy makers have started to consider the different effects that AI agents may have across different dimensions of our lives. However, the literature exploring the broader effects of human-agent interactions is still underdeveloped. In this paper, we review the problem of traffic modulation by autonomous vehicles (AVs) in mixed traffic flows and extrapolate the learnings to the different modes of interaction between humans and AVs to the pair humans-AI agents. In doing so, we propose a preliminary taxonomy of relational archetypes based on literature on Human-Computer Interaction (HCI) and AV-human interaction and tentatively explore how the resulting framework may lead to new questions regarding human-agent interactions. Our effort is aimed at strengthening existing bridges between these two research communities, which share similar traits: autonomy, fast adoption, high impact, and great potential for economic transformation. Building on previous analogies between AI Agents and AVs (e.g., regarding autonomy levels), we anticipate this paper to spark scholarly debate on the different types of impact that agents may have on our societies, while inviting other researchers to expand the scope of their comparative analysis regarding AI Agents.

## Introduction

AI Agents have been broadly characterised as autonomous systems capable of pursuing complex, long-term goals and taking direct action in virtual or real-world environments, typically by integrating a general-purpose AI (GPAI) model with additional "scaffolding" such as reasoning frameworks, memory, and tool access (Oueslati & Staes-Polet, 2025; Toner et al., 2024). Scholars differ in how they conceptualise and evaluate AI agents. Kasirzadeh and Gabriel (2025) emphasise intrinsic dimensions (autonomy, efficacy, goal complexity, and generality), whereas Chan et al. (2023) focus on external characteristics (underspecification, directness of impact, goal-directedness, and long-term planning). Progress on agentic AI, from feature release by frontier model providers to sector-specific adoption, cannot be understood without examining advances in General-Purpose AI (GPAI) models. Since 2010 (Sevilla et al., 2022), the AI frontier (Anderljung et al., 2023) has been advancing rapidly, creating a landscape of ever-evolving capabilities at increasingly higher levels of performance.

More recently, the AI frontier has shifted from models relying solely on training compute (like those released in late 2022 and early 2023) towards more balanced approaches that scale inference compute through reasoning and other post-training mechanisms (Ord, 2025). This shift, combined with advances in scaffolding techniques such as tool access, memory systems, and chain-of-thought reasoning] partially explains the boom in agentic AI (Kasirzadeh and Gabriel, 2025; Oueslati & Staes-Polet, 2025; Toner et al., 2024).

Beyond definitional debates and given that the central claims we make remain valid under different conceptions of AI agents, the broader concept agents allude to is posed to significantly transform society. As Tomasev et al. (2025) argue, current trajectories in Agent research and adoption may lead to a vast and highly permeable agent economy, where AI Agents operate in the real world at scales and frequencies beyond current capabilities for human oversight. The potential economic impact of AI Agents is substantial.

Autonomous Vehicles (AVs) underwent a similar phase, in which the potential of economic transformation received significant attention (McKinsey, 2025). Parallels between AVs and AI Agents extend beyond economic impact: from their relationship with AI to their autonomy levels, both technical objects can be productively compared for multiple purposes.

While this comparative potential is recognised in the still emerging contemporary literature on AI Agents, prior research has focused primarily on so-called levels of autonomy (Smakman et al., 2025, Feng et al., 2025; Morris et al., 2025). This work examines how autonomy can be conceptualised as a spectrum of user-agent relations, where responsibility and control gradually shift from human operators to developers as agents move from human-in-the-loop to fully independent operation, and how such gradations can help standardise evaluation, liability allocation, and risk management across increasingly agentic systems.

This article advances this research direction by proposing an expanded comparative analysis between AVs and AI Agents. Specifically, we present a preliminary framework to categorise different modes of interaction between AI Agents and humans, building on existing literature from AVs studies and Human-Computer Interaction (HCI). In Section 2 we introduce the problem of traffic modulation in AVs to explore how this subset of AV-human interaction problems has been characterised. In Section 3, through a preliminary review of HCI literature, we expand a tentative taxonomy of engagement forms into a proposed framework for characterising AI Agent-Human interactions. In Section 4 we discuss directions for future work.

## AV-Human Interaction

Over the past decade, many of the promises of AVs have materialised. From automated taxis and delivering services in some US cities (like Phoenix or San Francisco), to increasingly sophisticated driving aids (Lee et al., 2024), several companies have launched a diverse range of products and services where regulations permit. Whether AVs will fully supplant traditional vehicles or not remains an open question. However, mixed traffic (i.e. traffic flows composed of manually driven (MVs) and autonomous vehicles) in varying proportions will become increasingly common (Jo et al., 2024) as AVs adoption accelerates.

Mixed traffic has been an object of study for several years. Research examining the effects of AVs on overall traffic flow (Goldental & Kanter, 2020) and interaction patterns between AVs and MVs (Van Nunen et al., 2012; Liu, 2024; Huang et al., 2022; Jo et al., 2024) exemplify this body of work. In this regard, mixed traffic presents a particularly valuable research context because, as a transitional state (from fully manual to increasingly automated vehicle fleets), it offers enhanced visibility into friction points. The transition from MVs to AVs represents a further step in the social shift from the 'freedom of the open road' to an increased 'cocooning' (Wells & Xenias, 2015), while simultaneously providing concrete, tangible and actionable data into an ecosystem where high-agency technology interacts with humans.

Any social interaction in the road requires communication, coordination, and competition (Vanderhaegen et al., 2006) and is effectively governed by explicit rules (such as turn signals) and implicit ones (like body language) (Liu, 2024:1540). However, humans behind the driving wheel do not always display prosocial behaviours derived from rational and careful consideration: anger, aggression, road rage or vengeful driving characterise a portion of on-the-road interactions (Liu, 2024).

One particularly promising line of research in mixed traffic has been the study of traffic modulation, or how the introduction of AVs may affect traffic flow parameters like overall speed or congestion (Zheng et al., 2020; Wu et al., 2018). What makes AV-MV interaction (or mixed traffic) studies particularly valuable is that AVs, as an autonomous technology, provide real-world data to explore and think about how humans and increasingly agentic systems interact. Moreover, if the parallelism holds, ongoing work examining the AV-MV interactions could inform frameworks to better understand how humans interact with AI agents.

## A Comparative Analysis: AVs, Agents, and HCI

The Society for Automotive Engineers (SAE) proposes a six-level taxonomy (from 0 to 5) that classifies driving automation levels based on the breadth of tasks that are no longer executed by the human driver. This results in three main clusters: level 0 (No automation), levels 1 and 2 (assisted driving) and levels 3-5 (automated driving) (SAE, 2025). Consequently, the notion of an AV encompasses a broad range of products with varying autonomy and capability levels.

Similarly, AI agents have been characterised in many ways: from their functional attributes to their technical composition, agency remains a flexible concept permitting multiple manifestations. On the one hand and from a functional perspective, agents are tied to the capacity to autonomously pursue complex and underspecified goals, engage in long-term adaptive planning, and act within both virtual and real-world environments (Chan et al., 2024; Kapoor et al., 2024; Shavit et al., 2023). On the other hand, they technically constitute compound systems combining a general-purpose AI (GPAI) model, with "scaffolding" software that supports functions such as reasoning, memory, and tool use (Toner et al., 2024). But as the parallel with AVs shows (e.g., in Smarkman et al., 2025 or Feng et al., 2025) agency ultimately relates to reduced human involvement in executing a set of actions (i.e., to different degrees or levels of autonomy). In fact, and expanding the functional definition, AVs could be technically seen as a type of Agent.

In parallel, and before the AV and AI agent booms, extensive research on human-computer interaction (HCI) has examined questions equally relevant to both domains: namely, how humans interact with technology and how technology can be designed in ways to prompt new forms of interaction.

Drawing from HCI research, we can identify at least four ways in which humans may interact with AVs and,

analogously with AI Agents. These include cooperation, assistance, competition, and confrontation. Moreover, and building from the literature on AV and human interactions, these capture different modes of cooperation, competition, and assistance. In this context, cooperation among two or more actors can be broadly understood as the process by which actors pursuing conflicting goals act to facilitate all activities and reduce interference (Vanderhaeghen, 2006, p.2). An extreme mode of cooperation is assistance, in which the focus is on achieving a single actor's goals. Competition, on the contrary, can be conceptualised as a mode of interaction among two or more actors which seek to achieve goals that mutually interfere, and which leverage these interferences to enable their own success (Ibid.). An extreme mode of competition is confrontation, in which conflicting goals lead to a dispute. In this context, relational archetypes refer to distinct patterns of interaction that characterise how humans and autonomous systems (AI Agents, in this case) interact, defined by the degree and nature of mutual influence, goal alignment and control.

It is worth noting that the same interaction can relate to multiple modes, depending on the subset of actors considered and their relative goals. For instance, an AV may assist the person it drives, but at the same time, it may be in competition with a taxi driver whose livelihood the AV's activity undermines. As such, it seems plausible that the same human-AV interaction can simultaneously represent all four modes. Future work is required to determine the normative and analytical considerations for which actors to consider when classifying any given interaction, and the relative weight to attach to the goals of these actors.

In a survey of 291 papers, Shi et al. (2023) provide a taxonomy of human-GenAI interactions, aiming to offer a comprehensive overview of the building blocks constituting human-GenAI interactions (Shi et al., 2023, p. 2). While the article is focused on "GenAI", which is vaguely defined to reflect the multiple definitions in the literature, including those derived from seminal papers introducing key architectures such as GAN (Goodfellow et al., 2014) or transformers (Vashwani et al., 2017), it proves useful to examine agents as well. Even though Shi et al. focus on GenAI as systems, products, or models, the taxonomy ultimately addresses the core technology underpinning these objects, which also supports contemporary AI agents.

Through the lens of user-centered design, the authors propose several dimensions for analysing GenAI-human interaction. For this article's purpose, however, the most valuable dimension addresses the different levels of engagement, or how the interaction between GenAI and humans has been designed. This yields four primary modes of engagement:

- **passive**, which refers to an interaction where a human receives input from an AI without direct interaction.
- **deterministic**, or an interaction in which the outcome by the AI is determined by its inherent logic, instructions or rules, not shaped by the user.
- **assistive**, or an interaction whereby the AI supports the user without substantially contributing to the final product or in the creation process.
- **cooperative**, or an interaction in which both the AI and the user 'work together', collaboratively, to attain the final product. This is described as the most common mode of interaction at the moment. (Shi et al., 2023, pp. 18-9).

This taxonomy captures an important portion of the literature and provides a structure to categorise modes of engagement between GenAI and humans. However, a preliminary review of literature addressing AV-human interactions reveals a significant missing engagement mode: competition. As was established previously, any of the four modes between a human and AI depicted above can simultaneously be perceived as competitive by other humans (e.g., an AI Agent may "assist" a research department in conducting certain activities related to scientific discovery, such as form, refine or evaluate hypothesis (Jansen et al., 2024), but is at the same time competing with other humans who would otherwise execute the scientific tasks in the absence of an automated solution).

| Relational Archetype | Example |
|---|---|
| Passive | An AI Agent running in the background monitoring emails and filtering notifications. |
| Deterministic | An AI Agent that executes pre-defined workflows (e.g., creating periodic email backups). |
| Assistive | An AI Agent that provides administrative support summarises meetings or organises files. |
| Cooperative | An AI Agent that analyses large volumes of data and a human expert provides guidance. |
| Competitive | An AI coding agent that completes with junior developers (assisting the same senior engineers or project managers). |
| Confrontational | An AI Agent moderating content, deleting online comments against the will of the comments' human creator. |

Table 1: Overview of six relational archetypes applied to six different examples of an AI Agent.

From this, two preliminary conclusions emerge. First, competitive engagement, together with confrontation as its edge case, completes the set of agent-human interaction archetypes. Second, the same instance of an AI in general, and an agent specifically, interacts and affects multiple actor types concurrently which may pursue distinct goals, each one of these six modes can occur at the same time.

## Future Work

Having established a conceptual background, future work will include several efforts. Firstly, we aim to conduct empirical research to validate and understand how the proposed archetypes match real-world applications. To do so, systematic literature reviews or observational studies could be used to examine which modes are most prevalent, how users perceive and experience such archetypes, and how frequently and under what configurations do multiple archetypes co-occur.

Moreover, the parallel between AVs and AI Agents could be further extended. Current AVs are designed to be defensive. Given the complexity of human contexts and AV's inability to fully capture it, designers have defaulted to producing risk-averse and cautious AVs to prevent accidents. By extending the comparative analysis, and through the lens of relational archetypes, design objectives and conditions for AI Agents can be better characterised.